\documentclass[trackchanges]{aastex701}

\begin{document}

\title{Observational planning for the 2026 August 5 Falcon 9 Upper Stage lunar impact}

\author[orcid=0000-0002-7321-8401,sname='Fernando']{Benjamin Fernando}
\affiliation{Los Alamos National Laboratory, NM, USA}
\email[show]{benjaminf@lanl.gov}  

\author[orcid=0000-0003-3397-1682]{Jennifer Heldmann} 
\affiliation{NASA Ames Research Center, CA, USA}
\email{jennifer.heldmann@nasa.gov}

\author{Bill Grey}
\affiliation{Project Pluto, ME, USA}
\email{pluto@projectpluto.com}

\author[orcid=0000-0002-2294-119X]{John Ortiz} 
\affiliation{Los Alamos National Laboratory, NM, USA}
\email{jportiz@lanl.gov}

\author[orcid=0000-0002-5398-3456]{Bryan Euser} 
\affiliation{Los Alamos National Laboratory, NM, USA}
\email{beuser@lanl.gov}

\author[0000-0002-0726-6480]{Darryl Z. Seligman}
\affiliation{Department of Physics and Astronomy, Michigan State University, East Lansing, MI, USA}
\email{dzs@msu.edu}  

\author[orcid=0000-0002-0883-6748]{Eunhyeuk Kim} 
\affiliation{Korea Aerospace Research Institute, Daejeon, South Korea}
\email{eunhyeuk@kari.re.kr}

\author[orcid=0000-0002-5847-2241]{Anthony Colaprete} 
\affiliation{NASA Ames Research Center, CA, USA}
\email{anthony.colaprete-1@nasa.gov}

\author[0000-0001-6693-0014]{Elisa Maria Alessi} 
\affiliation{Istituto di Matematica Applicata e Tecnologie Informatiche, Consiglio Nazionale delle Ricerche, Italy}
\email{elisamaria.alessi@cnr.it}

\author[orcid=0000-0001-8690-3507]{Detlef Koschny} 
\affiliation{Technical University of Munich, Germany}
\email{detlef.koschny@tum.de}

\author[]{Anthony Cook} 
\affiliation{University of Aberystwyth, United Kingdom}
\email{atc@aber.ac.uk}

\author[orcid=0000-0003-1665-5709]{Joel Green} 
\affiliation{Space Telescope Science Institute, Baltimore, MD, USA}
\email{jgreen@stsci.edu}

\author[orcid=0000-0001-7150-2196]{Patrick King} 
\affiliation{Johns Hopkins University Applied Physics Laboratory,  Laurel, MD, USA}
\email{patrick.king@jhuapl.edu}

\author[]{Stacy Teng} 
\affiliation{Johns Hopkins University Applied Physics Laboratory,  Laurel, MD, USA}
\email{Stacy.Teng@jhuapl.edu}

\author[orcid=0000-0003-1582-0581]{Dawn Graninger} 
\affiliation{Johns Hopkins University Applied Physics Laboratory,  Laurel, MD, USA}
\email{Dawn.Graninger@jhuapl.edu}

\author[orcid=0000-0003-3489-8915]{Arnold Goldberg} 
\affiliation{Johns Hopkins University Applied Physics Laboratory,  Laurel, MD, USA}
\email{Arnold.Goldberg@jhuapl.edu}

\author[]{William Cooke} 
\affiliation{NASA Marshall Spaceflight Center, Huntsville, AL, USA}
\email{william.j.cooke@nasa.gov}

\author[orcid=0000-0001-8671-5901]{Mike F. Skrutskie} 
\affiliation{Department of Astronomy, University of Virginia, Charlottesville, VA}
\email{mfs4n@virginia.edu}

\author[orcid=0000-0001-5761-6779]{Kevin Schlaufman} 
\affiliation{Department of Physics \& Astronomy, Johns Hopkins University, MD, USA}
\email{nschmerr@umd.edu}

\author[orcid=0000-0002-3256-1262]{Nicholas Schmerr} 
\affiliation{University of Maryland, College Park, MD, USA}
\email{nschmerr@umd.edu}

\author[orcid=0000-0002-2332-3924,sname='Donahue']{Carly M. Donahue}
\affiliation{Los Alamos National Laboratory, NM, USA}
\email{cmd@lanl.gov}  

\author[orcid=0000-0002-6917-3458]{Carl A. Schmidt}
\affiliation{Boston University, Boston, MA, USA}
\email{schmidtc@bu.edu}  

\author[orcid=0000-0002-9984-4670]{Nancy J. Chanover}
\affiliation{New Mexico State University, Las Cruces, NM, USA}
\email{nchanove@nmsu.edu}  
 
\begin{abstract}

On 2026 August 5, at approximately 06:35 UT, a spent Falcon 9 upper stage will impact the lunar surface near Einstein Crater. This event will occur on sunlit terrain near the eastern limb as seen from Earth. The impact flash and resultant ejecta plume from this event are potentially observable from ground- and space-based observational facilities. This event provides an opportunity to attempt the recording of an artificial impact in real-time; although many of the properties of the event (such as visual magnitude) are imprecisely predicted at present. Moreover, this event provides an opportunity to test a pipeline for localising impacts on the lunar surface for future seismic experiments, investigating the dust and plume dynamics from impact events on the Moon, and considering hazards from artificial space debris impacts. Both professional and amateur astronomers are encouraged to attempt observations of this event.   

\end{abstract}

\keywords{\uat{Lunar impacts}{958} --- \uat{The Moon}{1692}}


\section{Introduction} 

Meteoroid impacts are commonplace on the Moon, and for the last few billion years have been the dominant lunar re-surfacing mechanism \citep{head2010global}. During the Apollo era, larger natural impacts (masses on the order of tens to hundreds of grams or more) were shown to be excellent seismic sources with a known (zero) depth, while smaller natural impacts were found to be the origin of the background ‘microseismic hum’ \citep{nakamura1982apollo}. These seismic measurements offer insight into the dynamics and properties of small-body impactors in the Solar System.

Artificial impacts on the Moon are much rarer, having occurred only a handful of times ever \citep{latham1970seismic}. Nonetheless, they are particularly valuable as probes of impact dynamics as they involve sources with a priori determined properties (mass, volume, and speed) and are generally anticipated (i.e. have an estimated occurrence time and location). Consequently, the uncertainties associated with using said flashes as ‘controlled sources’ to study impact cratering dynamics or seismic wave propagation are much reduced because the source properties are independently known \citep{nunn2024artificial}. Furthermore, observation campaigns can be planned in advance to enable optimised data gathering during the event, which is generally not the case with natural impacts \citep{fernando2022seismic}.

\subsection{Past artificial impacts on the Moon} 

The first artificial impact on the lunar surface occurred in 1959, when the USSR’s Luna 2 probe deliberately collided with the Moon. In 1963, it was reported (via the Royal Aircraft Establishment in the UK and then via NASA) that Soviet astronomers had observed the impact flash produced by Luna 2's impact, but this report has never been confirmed with images or other evidence \citep{gehring1963investigation}. A possible dust plume associated with the Luna 5 impact in 1965 was also potentially captured in the USSR, but was not revealed by the authorities there at the time \citep{ksanfomality2018luna}. 

Of US lunar missions, a number subsequently included spent spacecraft parts (Saturn IV-B upper stages and lunar modules) which had planned impacts during the Apollo era (1969-72). These acted as controlled sources for the Apollo Passive Seismic Experiment (PSE), though no confirmed impact flashes from these events were recorded in this era before modern camera technology was available. Nonetheless, these seismic data continue to prove incredibly valuable in studies of the lunar interior \citep{latham1970seismic}. 

Following Apollo, Japan's Hiten spacecraft was de-orbited into the Moon in 1993, followed by ESA's SMART-1 spacecraft in 2006 \citep{cudnik2009lunar,veillet2007smart, Burchell2010SMART1LunarImpact}. Both spacecraft were likely immediately vapourised on impact \citep{Burchell2010SMART1LunarImpact}.

In 2009, NASA's LCROSS mission involved deliberately impacting a Centaur rocket upper stage into the lunar south pole region at Caebus Crater to produce an impact plume \citep{heldmann2012lcross}. The Centaur of 2,305~kg mass impacted a permanently shadowed region at approximately 2.5 km/s. A shepherding spacecraft then characterised the plume via a set of nine instruments to measure the dynamics and composition. The second spacecraft impacted the lunar surface a few minutes later \citep{colaprete2012overview}. NASA's Lunar Reconnaissance Orbiter also imaged the plume spectroscopically to determine its composition \citep{gladstone2010lro}, and the sodium in the plume was spectroscopically detected from the Earth \citep{killen2010observations}. 

LCROSS met its primary aim of identifying the composition of the regolith and material contained within it at the lunar south pole. The concentration of water ice was found to be approximately 5.6\% \citep{colaprete2010detection}. Subsequent imaging of the crater was challenging due to its location in a permanently shadowed region, but recent measurements indicate a 22~m diameter crater with an estimated $\sim$350~metric tons of material excavated \citep{fassett2024lcross}. 

Determining the exact crater shape of the LCROSS crater remains  challenging, but it should be noted that deviations from a circular footprint for natural impactors are only expected for impacts at very shallow angles (a few tens of degrees from the horizontal, \cite{elbeshausen2013transition}). For artificial impactors, which are generally slower, irregularly shaped, and underdense (somewhat hollow), more complex, non-circular craters may result \citep{rajvsic2021numerical}. For example, the 2022 impact of China's Chang'e 5 lander upper stage occurred in a non-polar area, though it occurred on the lunar farside and hence was not observed in real time. Chang'e 5's upper stage produced a double crater, likely due to the `decapitation' of the impactor. Furthermore, any unspent fuel in rocket stages may explode upon impact, resulting in greater energy release than from a natural object of equivalent momentum. 

\subsection{Telescopic observations of impacts}

During the last two Apollo missions, astronauts in the orbiting command module reported multiple impact flashes believed to have been caused by natural meteoroids striking the lunar surface \citep{fernando2026cross}. Since 1999, several hundred similar impacts have been recorded telescopically from Earth \citep{bellot1998observation, suggs2015results, liakos2024neliota}. The resulting impact flashes are only visible on the darkened (non-sunlit) hemisphere of the Moon. The brightest have visual magnitudes around +3 and the current night-side detection limit is around +11 for the most sensitive instruments. The flashes themselves last less than a second, necessitating high-cadence imaging if the light curve is to be resolved. Although there is some uncertainty about the fraction of the impactor’s energy which is converted into light, and variability depending on the target material, these flashes are generally assumed to have been caused by impactors in the sub-kilogram range moving at velocities of tens of kilometres per second. In addition to measurements of flash timing and location, recent studies have also been able to resolve the temperature of some flashes via multi-channel detections \citep{madiedo2018first,az612neliota, avdellidou2019temperatures}.

In addition to the flash, it is also expected that these impacts will produce a plume of ejecta (and potentially vapour as well). The dynamics of impact plumes are complex, involving both emission at both high and low angles to the horizontal \citep{bernardoni2019impact}. Because natural impacts cannot be predicted in advance to any meaningful degree (aside from knowing that the mean impact rate will be elevated during meteor showers), they are only fortuitously captured \citep{berezhnoy2019detection}.

Fortunately this is not the case for many artificial impacts, where in general prior knowledge of the impact time and location means that optimised observing campaigns can take place. The impact flash from Hiten was captured by the Anglo-Australian telescope \citep{cudnik2009lunar} in 1993, and SMART-1's impact in 2006 was resolved using the 4-metre class Canada-France-Hawaii Telescope \citep{veillet2007smart}. However, the magnitude of this flash was never robustly determined, meaning that no actual measurement of flash magnitude from an artificial lunar impact has ever been published \citep{Burchell2010SMART1LunarImpact}. 

During the LCROSS impact in 2009, a number of observatories attempted to observe the upper stage collision. Unfortunately, the impact flash was not directly observable from Earth due to obscuration by small hill on the lunar surface. As such, the focus from Earth was on observing the plume itself to determine its dynamics (e.g size, altitude, optical depth, and settling time) and potential water content. 

Both ground-based (e.g. Apache Point, New Mexico) and space based observatories (Hubble Space Telescope, Lunar Reconnaissance Orbiter) attempted to observe the impact (e.g. \citep{chanover2011results, heldmann2012lcross}). Ground based observatories initially reported uniform non-detections of the plume. However, with more modern and advanced processing techniques, evidence of the plume from LCROSS's upper stage in previously collected data has emerged \citep{strycker2013characterization} with an optical depth of 0.0018$\pm$0.0002 (comparable to the shepherding spacecraft's measurements of 0.002 to 0.003, though in a different band). A tentative detection of OH, possibly from the plume, was also reported in Hubble data \citep{storrs2010observations}. 

The 2009 impact flash of the SELENE (Kaguya) spacecraft operated by JAXA was also observed from at least two professional observatories in India and Australia \citep{shirao2011kaguya}. This impact occurred very close to the terminator, and hence bright scattered light from the lunar dayside made it challenging to observe. Bad weather across much of East Asia prevented amateur observers from recording the impact, though such a campaign was attempted. 

For missions other than SMART-1, LCROSS, and Kaguya which impacted the lunar surface at high speed (e.g. Lunar Prospector, GRAIL), no impact flash or plume detection has been reported. It is also unclear whether any deliberate effort was made to observe them, despite GRAIL's impact being on the lunar nearside. 

For completeness, we also note that both professional and amateur observations of artificial impacts on other small bodies have also been made, in the professional case from the ground as well as from space. These include observations of comet 9P/Tempel 1 during the Deep Impact mission and asteroid moonlet Dimorphos during the DART mission. These observations assisted with determining the ejecta mass, the ratio of dust to gas within said ejecta, and the increase in activity of the comet in the case of Tempel 1 (e.g., \citet{meech2005deep, lisse2006spitzer, graykowski2023light, li2023ejecta, weaver2024lucy}). 

\section{The 2026 August 5 Falcon 9 Upper Stage impact}

\subsection{Background}
In May 2026, it was noted that the spent upper stage of a SpaceX Falcon 9 rocket was expected to impact the lunar surface on the (UT) morning of 2026 August 5 \citep{gray2026upperstage} at 06:35~UT (JD 2461257.774). This piece of space debris (COSPAR 2025-010D, NORAD/SatCat 62719) originally boosted Firefly’s Blue Ghost-1 and the ispace Hakuto-R Resilience landers on their way to the Moon in 2025 and was abandoned in a Moon-crossing high-Earth orbit. 

\subsection{Impact particulars}
\label{sec:ip}
\begin{figure*}
\begin{center}
    \includegraphics[width=0.5\linewidth]{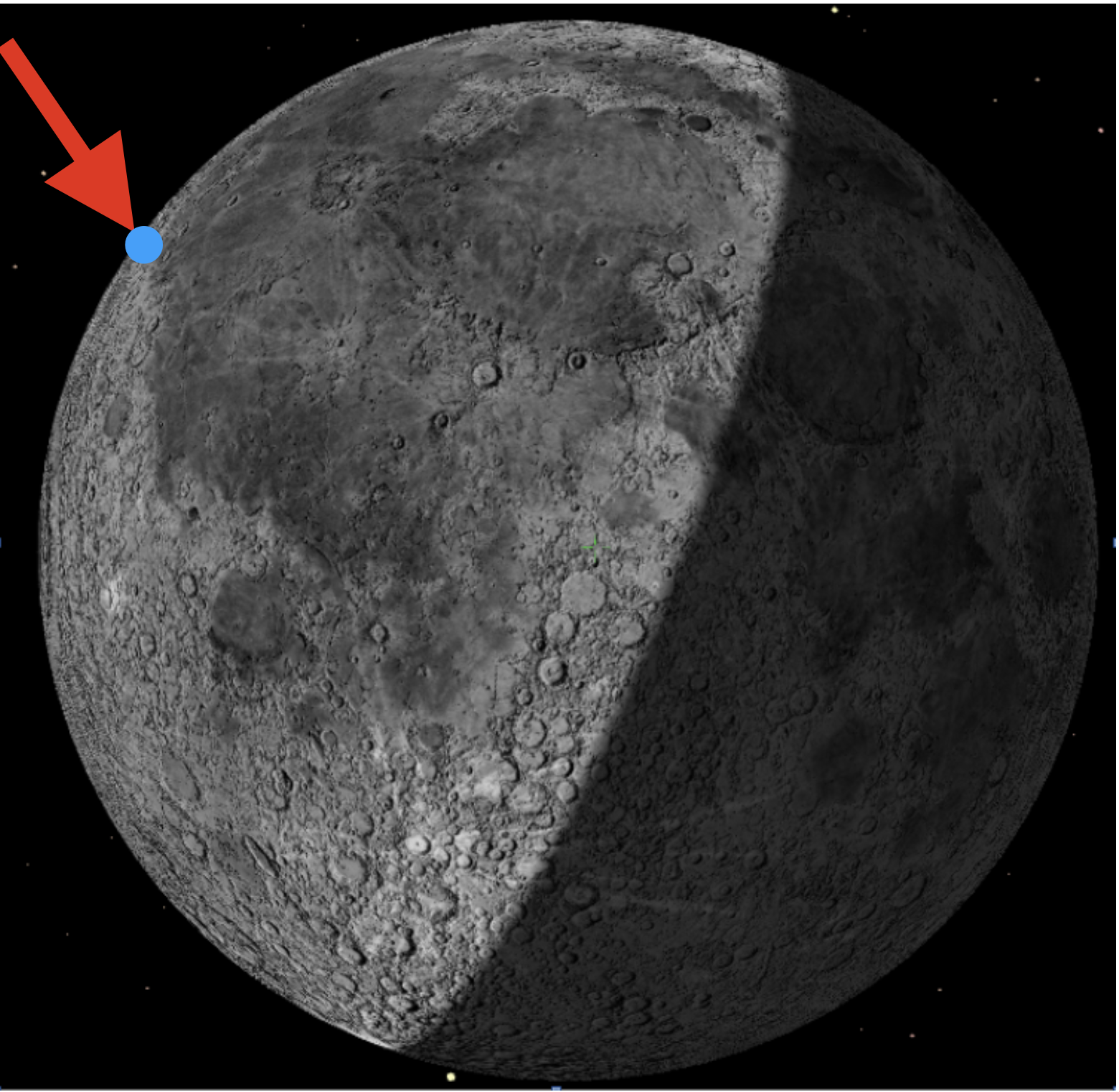}
    \caption{The estimated impact location of the Falcon 9 upper stage on the lunar surface is shown with a blue dot. On 2026 August 5 the Moon will be near last quarter, around 56\% illuminated. This figure accounts for the effects of lunar libration, without which the impact would occur over the edge of the lunar limb. Image modified from \citep{gray2026upperstage} with permission.}
    \label{fig:onlyfig}
    \end{center}
\end{figure*}

The expected impact speed is 2.43~km/s at an angle of 34$^{\circ}$ from the vertical in the vicinity of Einstein Crater (88$^{\circ}$W, 15$^{\circ}$N). The exact orientation of the spacecraft body at the time of impact is difficult to predict  \citep{gray2026upperstage}. The estimated position of impact, along with the lunar phase at the time of impact, is shown in Figure \ref{fig:onlyfig}. Immediately prior to impact, the upper stage is expected to undergo a close conjunction (on the order of kilometres) with the Korean Pathfinder Lunar Orbiter spacecraft. The Lunar Reconnaissance Orbiter will be orbiting at a lunar longitude that is nearly orthogonal to the impact site, which unfortunately precludes ultraviolet spectroscopy of plume gases \citep{gladstone2010lro}. 

The exact mass of the spacecraft is not known, but it is expected to be approximately 4,000~kg (assuming all propellant is spent); yielding a kinetic energy at impact of 11.8~GJ and a momentum of 9.7~MNs of which 5.5~MNs is vertical with respect to the lunar surface normal immediately prior to the point of impact. The upper stage's length and diameter are approximately 12~m and 4~m, respectively. This object is significantly more massive than most natural lunar impactors, but is also moving much more slowly. Modelling and laboratory studies indicate that flash brightness decreases nonlinearly with impact velocity, making slow space debris impacts disproportionately dim as compared to faster natural impactors. This is because the visual magnitude of the flash decreases significantly once the impactor's velocity is lower than the speed of sound in the target material \citep{ernst2002effect, swift2010exponential} --- natural meteoroids all produce supersonic (shock)waves in all lunar materials, but space debris moving at $\sim$2~km/s will only produce a shockwave if the impact is into regolith (sound speed $\sim$ hundreds of metres per second) rather than bedrock (1000-2000~m/s). This produces uncertainty that makes the visual magnitude of the flash challenging to predict because the current impact ellipse also includes bedrock exposures.

\subsection{Potential observables}

We consider three distinct phenomena resulting from the impact, and their potential observability: 

\begin{enumerate}
    \item The impact flash: a brief flash of light lasting $<1$~s in duration produced by the vapourisation of the impactor,
    \item The impact plume: a cloud of ejecta (and possibly vapour) lofted above the lunar surface and evolving on a timescale of minutes, and
    \item The resultant crater: a final crater diameter of 20-30~m is expected based on Chang'e 5 and LCROSS data, and modelling. A double crater may be produced if the impactor is decapitated, which may occur depending on the stage's attitude at the time of impact. 
\end{enumerate}

Each of these phenomena will require different tools to observe. 

\subsubsection{The Impact Flash}

Detection of the impact flash is expected to be challenging due to its location on the dayside hemisphere of the Moon. There has not been a natural or artificial impact flash detected on the lit portion of the lunar surface to date. 

The visual magnitude, $M$, of a flash is highly dependent on both  the orientation of the upper stage at impact and the target material (regolith versus bedrock). Modelling results for the flash brightness range from $M = +3$ at the brightest to undetectably dim (fainter than $M = +15$ for an impactor hitting bedrock), largely due to the variation in the assumed luminous efficiency because of the unknown regolith versus bedrock target properties (e.g. \citet{ernst2002effect}). This wide range of predictions highlights the need to validate these models against observable lunar flashes for slow-moving impactors, which this event offers an opportunity to achieve. 
    
If any observations are to be attempted, they will require a telescope rather than binocular or naked-eye measurements. Observing may take place in visible or infrared bands, with their relative merits discussed below. High-cadence imaging (ideally at least 20+ frames per second) would be optimal for the flash, such that the light curve can be resolved and the flash is not missed in the time between frames (lower cadence observations would be sufficient for detecting the plume).

With the enormous variation in predicted brightness it is difficult to meaningfully advise as to the required telescope aperture. However, it is useful to note that past impact flash observing campaigns have involved telescopes with apertures as small as 4" (10~cm, \citet{cudnik2003ground}) and in general expert observers do not dissuade those with small apertures from trying to record flashes \citep{cudnik2009lunar}.

Observations of impact flashes have been made routinely in V, R, and I bands \citep{madiedo2018first, avdellidou2019temperatures, liakos2024neliota}, and recent results have also suggested that the J band may be advantageous for observers seeking to record flashes in daylight due to the reduced sky brightness, which for this event would include observers in Europe and Africa \citep{sheward2024extending}. Narrower isolation of particular emission lines may also be fruitful, for example lithium contamination from Falcon 9 upper stages entering the Earth's atmosphere has been observed before \citep{wing2026measurement} and lithium is known to have a strong and accessible doublet emission line (Li I) at 670.8~nm. However, inability to use one of these relatively expensive specialised filters should not dissuade observers (especially the amateur community), as even unfiltered observations are potentially valuable. 

The most useful data products to report would be the flash location, timing, and light curve; though even timing and location alone are still of scientific value. Even if the flash is not immediately obvious in recorded images, differencing (subtracting the image taken at the flash time from a higher-resolution stack taken immediately before or after it) may reveal it. Manual observations through a telescope (i.e. without a camera) are less likely to result in success but may still be attempted to confirm flash timing. 

We re-emphasise that the magnitude of this impact flash is not precisely known, and at the dimmer magnitude estimate it is possible that observers even with large telescopes may not record anything. Nonetheless, given that attempting the observations is reasonably straightforward, we encourage interested parties to take part.

\subsubsection{The impact plume} 

The dynamics of an impact plume are complex, with both high-angle and low-angle contributions. Using the Hybrid Optimization Software Suite (HOSS) multiphysics simulator \citep{knight2020hoss} we performed a nonlinear shock physics simulation of the upper stage impact into lunar regolith to try and predict these dynamics. 

The HOSS simulation uses the finite-discrete element method (FDEM), which combines aspects of discrete- and finite-element methods. The simulation consists initially of continuous, solid domains discretized into finite elements. The individual solid domains can interact with one another and develop new discontinuities or undergo fragmentation upon meeting some failure criterion, thereby creating new discrete domains. HOSS has been used to simulate a diverse array of problems, including high strain-rate, hypervelocity planetary impact problems \citep{froment2020lagrangian,froment2024numerical,caldwell2021benchmarking}. Note that this simulation does not account for the tiny fraction of the impactor's kinetic energy that is converted into electromagnetic radiation, and hence cannot be used to directly inform the measurements of the flash brightness discussed above. However, in terms of bulk energetics this is a justifiable assumption given that the fraction converted is negligible, on the order $10^{-3}$ of the total energy for natural meteoroids \citep{bellot2000luminous} and potentially lower for slower artificial impactors \citep{ernst2002effect}.

The properties of the impactor are simulated as described in Sec. \ref{sec:ip}, and the regolith is modelled as having a density of approximately 1,590~kg/m$^3$. At a resolution of 5 cells per projectile radius, the minimum resolved scale in our simulation is 0.37~m. Fragments of ejecta smaller than this are therefore not resolved, but this resolution is sufficient for our main objectives of placing a lower limit on the volume, mass, and vertical extent of ejected material. 

Due to the time-sensitive nature of this event and the computationally intensive nature of the simulation, we restrict our simulation to an axisymmetric, vertical, end-on impact (i.e. with the long axis of the impactor perpendicular to the surface). This is an end-member case, but is nonetheless instructive. The simulation is allowed to run for 0.90~s. This is sufficient for a transient crater to form, but not to reach a final steady-state. For completeness, we also compare the predictions from our simulation to those derived from crater scaling laws (e.g. Pi-scaling from \citet{housen2011ejecta}).

\begin{figure*}
    \centering
    \includegraphics[width=1.0\linewidth]{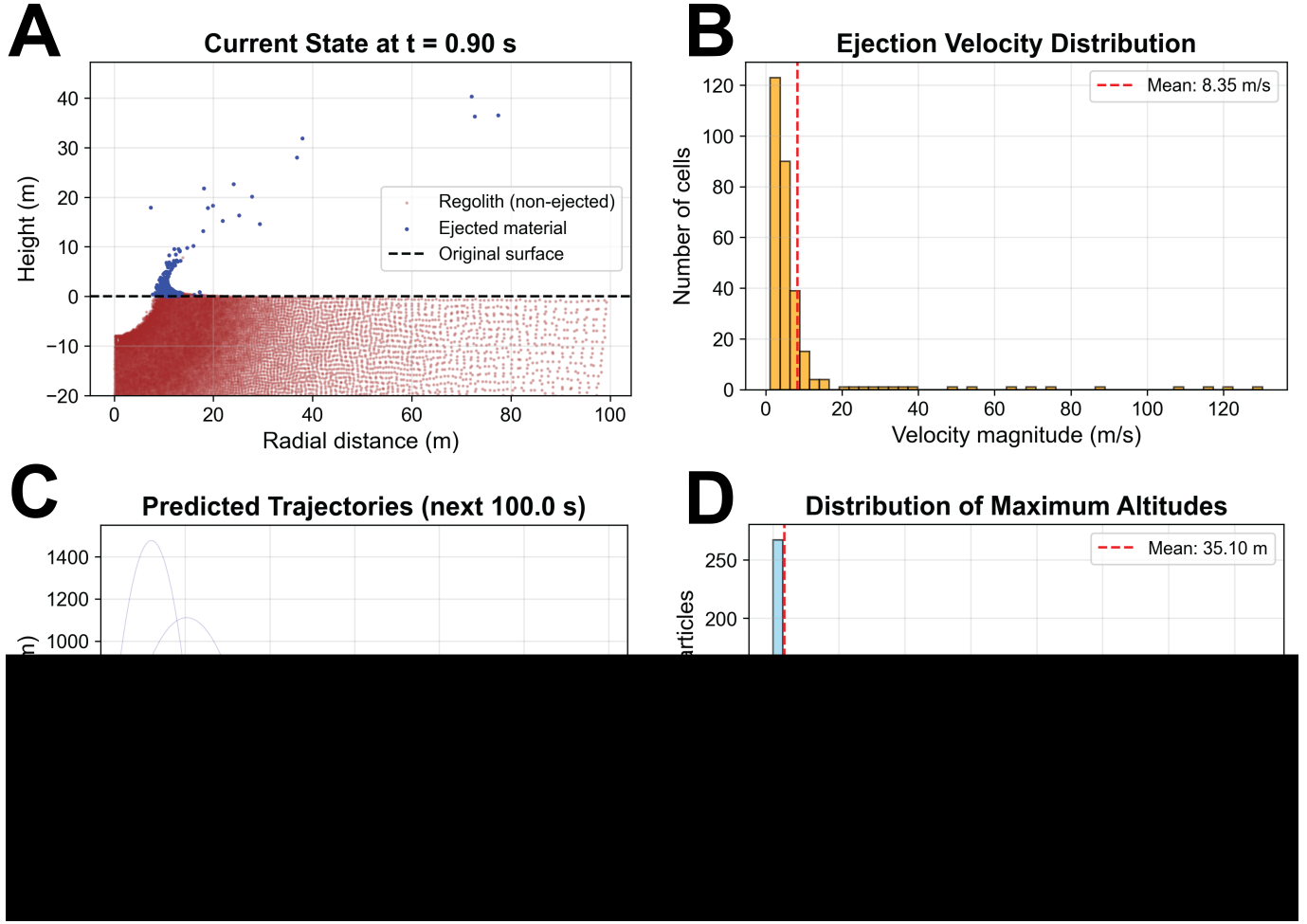}
    \caption{Results from the HOSS simulation of the Falcon 9 upper stage impact assuming an end-on (long axis perpendicular to the surface) attitude. \textbf{A} Snapshot of the impact region at the end of the simulation (0.90~s). Excavated material is coloured blue, while unexcavated material is coloured red. At this point in time, most of the excavated material is in the form of a transient `ejecta curtain', with higher-velocity particles having already left the simulation domain. \textbf{B} Distribution of ejecta particle velocities and \textbf{C} their resultant ballistic tracks, with the distribution of maximum altitudes shown in \textbf{D}. 
    }
    \label{fig:sims}
\end{figure*}

A summary of the results from this simulation are shown in Figure \ref{fig:sims}. The excavated lunar regolith material is defined as that which (i) is above the initial surface level of the target material following the impact and (ii) has a vertical (axial) velocity component $v_y > 1.0$ m/s. 

While this simulation tracks the majority of the excavated mass, the smallest (and most rapidly ejected) particles are not resolved. The smallest debris fragments are the most numerous, and contribute the most significantly to the total fragment surface area contained within the ejecta (which will affect the scattering of sunlight by the plume), but simply cannot be resolved in this setup. 

The velocity distribution follows the theoretical $v^{-3}$ distribution (a consequence of the scale-free distribution of fragment properties), meaning that the majority of particles are slow moving compared to a very fast-moving ejecta represented by the tail of the distribution. This can be seen in the skewness of both the velocity Figure \ref{fig:sims} Panel B and maximum altitude distributions Figure \ref{fig:sims} Panel D. 

The excavated mass from the crater determined via simulation is $\sim$1,120,000~kg. This is slightly less than the Pi-Scaling prediction based on impact scaling laws (e.g. \citet{housen2011ejecta}) of 1,200,000~kg. Both values indicate that the impactor will likely excavate around 150-200 times its own mass of regolith. 

Pi-scaling predicts a maximum ejecta velocity of 1,200~m/s. At the end (0.90~s after impact) of the HOSS simulation, the fastest ejecta particles in the simulation box are moving only 130~m/s. This is likely because the minimum fragment size resolved in the HOSS simulation is on the order of 0.4 m, and in reality it is expected that smaller fragments and entrained debris will have much higher velocities. 

Approximately 50\% of the particles resolved by the simulation re-impact the surface within 5~s (likely forming secondary craters), whereas only 2\% remain in flight after 30~s. Again, this means that the plume will be dominated by the long-lived ejecta; the longest-lived particles resolved in our simulation will remain ballistically lofted for $\sim$minutes, and hence smaller unresolved particles which are ejected at higher initial velocities may last for on the order of ten minutes.   

Vertically, the highest resolved particles in the HOSS model reach altitudes of $\sim$1.5~km. As noted above, smaller fragments which are not resolved in our simulation almost certainly reach higher. This means that the plume will almost certainly reach the few kilometres in height to be theoretically visible from Earth over the limb (as opposed to silhouetted against the illuminated lunar surface). We note that our simplifying assumption of a vertical impact likely overpredicts the maximum height of the plume compared to the expected oblique impact, but equally likely underpredicts the plume duration. 

Neither HOSS modelling nor Pi-scaling are able to directly predict the optical depth ($\tau$) of the plume. Scattering is likely to be dominated by the smallest and longest-lived particles. Assuming a plume similar to that produced by LCROSS, $\tau \sim$ 0.001 is a reasonable estimate \citep{strycker2013characterization}. We note that the plume's shadow is highly unlikely to be visible as the Sun will be nearly overhead the impact location when it occurs.  

Different observational strategies are possible to capture the plume, ranging from simple high-cadence imaging to track plume shape and brightness over minutes to spectroscopic measurements to measure composition (noting that little to no sub-surface water ice is expected at these latitudes). Given the expected impact location, the plume may potentially be detected by masking the entire lunar limb (acting as a `selenograph') or else excluding the lunar surface itself entirely from view. Ground-based observers should note that the sky will still be appreciably bright this close to the lit hemisphere, however. 

\subsubsection{The resultant crater}
    
The resultant crater will be too small to be directly resolvable from Earth, and will only be detectable by orbiting spacecraft with high-resolution imaging capability. Pi-scaling indicates that a crater diameter around 27~m is expected with a depth of around 5~m. The HOSS simulation does not reach a steady state and hence final crater dimensions cannot be extracted. 

Laterally, modelling suggests that the ejecta blanket should be continuous over an area around 70~m in diameter centred on the crater. This presents a large target region in which fresh material will be excavated and may be imaged by subsequent spacecraft flyovers. This will be particularly advantageous for understanding the composition of the target material. The maximum ballistic range of resolved particles is found to be slightly under 1,000~km, which is far enough to be significant from the perspective of presenting a hazard to future astronauts or infrastructure across much of the lunar surface.

A number of spacecraft follow-up observations are planned. NASA's Lunar Reconnaissance Orbiter (LRO, \citet{robinson2010lunar}) will acquire both baseline (before) and post-impact (after) imagery.

Korea's Pathfinder Lunar Orbiter \citep{jeon2024korea} will experience a close conjunction with the rocket stage (on the order of a few kilometres or less) approximately two minutes prior to the stage's impact. Baseline imaging has already been acquired and follow-up is planned in as close to real time as possible (on the same UTC day). 

\subsection{Observing from the ground}

Observation time has been allocated on a number of ground-based professional observatories including high cadence imaging at the Astrophysical Research Consortium 3.5m telescope at Apache Point Observatory, spectroscopic observations of sodium (Na) at the 4.3~m Lowell Discovery Telescope, and spectroscopy targeting OH, Na and K emissions from the UVES instrument on the 8.2~m Unit Telescope 2 (UT2) of the Very Large Telescope. Observations are also being coordinated on an amateur level via the NASA-affiliated `Impact Flash!' citizen science project (\url{https://science.nasa.gov/citizen-science/impact-flash/}) and the Lunar Impact Flash Network (\url{https://lif.mi.imati.cnr.it}). 

For ground-based observers, two additional constraints exist: 

\begin{enumerate}
    \item Darkness: to have the best chance of detecting the impact, ideally an observer will be in local darkness. With an expected impact time of 06:35~UTC, this results in a geographical limit where observers will be in darkness in South America and low-to-mid latitude North America. Observing during daylight hours is in theory possible, but is generally more challenging due to the sky's brightness. Observations during daylight have been found to be optimised when using the J infrared band \citep{sheward2024extending}. 
    \item Lunar position: the Moon must necessarily be above the local horizon at the time of impact for ground-based observers. Within the Americas, this requirement excludes observers in Hawaii and Alaska. As seeing is better when the Moon is higher in the sky, conditions will favour observers in North America away from the West Coast. 
\end{enumerate}

\subsection{Preparation}

Where possible, we encourage observers to undertake practice runs the prior to the impact. This preparation enables observing strategies to be optimised and equipment tested. Conducting a practice run the day prior (UTC night of August 4) will enable testing under conditions that closely mimic the lunar phase illumination expected at the time of the event. 

For members of the amateur community interested in attempting to observe this event but who have not made flash observations previously, several `how-to' guides have been published which contain information on how to optimise telescope and camera setups (e.g. \citet{cudnik2009lunar} or \url{https://britastro.org/section_information_/lunar-section-overview/lunar-section-observation-activities/lunar-geological-change-detection/observing-lunar-impact-flashes}). 

\section{Data sharing}

Observers recording image data and who are interested in sharing their results are welcome to submit recordings through the Lunar Impact Flash Portal (\url{https://lif.mi.imati.cnr.it/lif.html}). Those making more complex measurements (e.g. spectroscopy) who are interested in collaborative investigations are encouraged to contact this team for more details. 

\section{Conclusions}

The upcoming Falcon 9 upper stage impact on the Moon on 2026 August 5, presents an ideal opportunity for observers in the Americas and space-based assets to study a significant lunar impact in real time. Potential observables from this event include the flash at the time of impact (likely lasting less than a second) and the ejecta plume (lasting minutes to tens of minutes) Although some specifics of the impact (such as its precise visual magnitude) remain unclear due to the complexity of the physical processes involved, this event is an ideal opportunity to calibrate measurements of flash brightness and plume dynamics against a known event, should either be detected.

This event also provides an opportunity to test pipelines for measuring flash properties to locate impact events seismically, and to better understand the multi-modal hazards posed to future lunar infrastructure and astronauts from space debris impacting the Moon.  It also provides an opportunity to test our capability to monitor the cislunar environment. Given the renewed interest in the lunar exploration, the impact of artificial bodies on the Moon could is almost certain to become more frequent. To avoid uncontrolled increases in the amount of space debris in lunar orbit (as is the case now in low Earth orbit), advancing monitoring of cislunar space from ground- and space-based facilities will be key. 

\begin{acknowledgments}

This manuscript has Los Alamos National Laboratory release number LA-UR-26-25004. 

BF is funded by the Frederick Reines Distinguished Fellowship at the Los Alamos National Laboratory. DZS acknowledges funding support from JWST GO 5959, which was provided by NASA through a grant from the Space Telescope Science Institute. NS and the Impact Flash! citizen science project acknowledges support from NASA SSERVI GEODES grant 80NSSC19M0216.EMA acknowledges support by the Italian Space Agency through the agreement n. 2024-6-HH.0, `Supporto scientifico alla missione LUMIO'. PK, ST, AG, and DG acknowledge internal funding support from the Space Formulation Mission Area of JHU/APL.

\end{acknowledgments}

\begin{contribution}

Conceptualisation - BF, JH, AC, BG, JG. KS;
Investigation - BF, JH, EMA, DK, AC, CMD, JO, BE, EK, CAS, NJC, PK, ST, DG, AG, MFS;
Writing: original draft - BF;
Writing - review and editing - BF, DZS, NS, WC.

\end{contribution}

%

\software{Project Pluto \citep{gray2026upperstage}, HOSS \citep{knight2020hoss}
          }


\bibliography{sample701}{}
\bibliographystyle{aasjournalv7}



\end{document}